\documentclass{sig-alternate-2013}
\usepackage{multirow}
\usepackage{flushend}
\def\sharedaffiliation{
\end{tabular}
\begin{tabular}{c}}

\newfont{\mycrnotice}{ptmr8t at 7pt}
\newfont{\myconfname}{ptmri8t at 7pt}
\let\crnotice\mycrnotice%
\let\confname\myconfname%

\permission{Permission to make digital or hard copies of all or part of this work for personal or classroom use is granted without fee provided that copies are not made or distributed for profit or commercial advantage and that copies bear this notice and the full citation on the first page. Copyrights for components of this work owned by others than the author(s) must be honored. Abstracting with credit is permitted. To copy otherwise, or republish, to post on servers or to redistribute to lists, requires prior specific permission and/or a fee. Request permissions from permissions@acm.org.}
\conferenceinfo{WebSci'14,}{June 23--26, 2014, Bloomington, IN, USA. \\
{\mycrnotice{Copyright is held by the owner/author(s). Publication rights licensed to ACM.}}}
\copyrightetc{ACM \the\acmcopyr}
\crdata{978-1-4503-2622-3/14/06\ ...\$15.00.\\
http://dx.doi.org/10.1145/2615569.2615699 }

\begin{document}

\title{
Evolution of Online User Behavior During a Social Upheaval}

\numberofauthors{1}
\author{
\alignauthor Onur Varol, Emilio Ferrara\titlenote{Corresponding author: \texttt{ferrarae@indiana.edu}}, Christine L. Ogan, Filippo Menczer, Alessandro Flammini 
\sharedaffiliation
\affaddr{Center for Complex Networks and Systems Research}\\
\affaddr{School of Informatics and Computing, Indiana University, Bloomington, USA}
}


\maketitle

\begin{abstract}
Social media represent powerful tools of mass communication and information diffusion. They played a pivotal role during recent social uprisings and political mobilizations across the world. Here we present a study of the Gezi Park movement in Turkey through the lens of Twitter. We analyze over 2.3 million tweets produced during the 25 days of protest occurred between May and June 2013. We first characterize the spatio-temporal nature of the conversation about the Gezi Park demonstrations, showing that similarity in trends of discussion mirrors geographic cues. We then describe the characteristics of the users involved in this conversation and what roles they played. We study how roles and individual influence evolved during the period of the upheaval. This analysis reveals that the conversation becomes more democratic as events unfold, with a redistribution of influence over time in the user population. We conclude by observing how the online and offline worlds are tightly intertwined, showing that exogenous events, such as political speeches or police actions, affect social media conversations and trigger changes in individual behavior.
\end{abstract}

\category{}{Human-centered computing}{Collaborative and social computing}[Social media]
\category{}{Information systems}{World Wide Web}[Social networks]
\category{}{Networks}{Network types}[Social media networks]


\section{Introduction}

Technologically mediated communication systems, like social media platforms and online social networks, support information sharing and foster the connectivity of hundreds of millions of users across the world every day \cite{boyd2010social,vespignani2009predicting}. 
The adoption of these platforms has been associated with profound changes in 21st-century society: they affect how we produce and consume information \cite{aral2012identifying,bond201261}, shifting the paradigm from a broadcasting model (one-to-many, like radio and TV) to a peer-to-peer (many-to-many) distribution system.
They have also altered the ways we seek information to understand societal events surrounding us \cite{metaxas2013rise,metaxas2012social}, and how we interact with our peers \cite{centola2010spread,centola2011experimental}. 

The use of online social media to discuss politics and policy has recently been associated with political uprisings and social protests around the world.
Prominent examples include the revolution in Egypt \cite{choudhary2012social}, the American anti-capitalist Occupy Wall Street movement \cite{byrne2012occupy,chomsky2012occupy,conover2013geospatial,conover2013digital}, and the Spanish May 15th protests \cite{borge2011structural,gonzalez2011dynamics}. Social media have played a pivotal role in the development and increasing frequency of these social movements \cite{bennett2003communicating,garrett2006protest,myers1994communication}. Using survey methodology, Tufekci and Wilson \cite{Tufekci2012} found that the use of social media in the Egyptian protests allowed people to make informed decisions about participation in the movement, provided new sources of information outside of the regime's control, and increased the odds that people participated in the protests on the first day.  Another survey found Facebook use for news and socializing in Chile's youth  movement to be  positively associated with participation in the protests \cite{Valenzuela2012}.  
Chief among social platforms used for protests is Twitter that, with more than a half billion users, provides a high-visibility window on real-world events and an active forum for discussion of political and social issues. The mostly ungoverned nature of this platform ensures a democratic, peer-to-peer discussion, aiming at both creating a framing language to set goals for the protest, and as a vehicle for mobilizing resources and social capital to sustain it \cite{aday10-1,howard2011opening,lotan_revolution,conover2013geospatial}.
Individuals and organizations can discuss and share information on Twitter about the movement's political and social objectives \cite{benford1997insider,benford2000framing}. They can also coordinate to marshal the resources needed to carry out on-the-ground activities like encampments or marches \cite{jenkins1983resource,mccarthy1977resource}. 

\begin{table*}[t]
\caption{List of relevant events during the protest divided in three categories.}
\centering
	\begin{tabular}{lccp{12cm}} 
	\hline
	& Code & Event date  & Event description \\
	\hline
	\multirow{4}{*}{\rotatebox[origin=c]{90}{Government}}
		& A1 & 2013-05-29 & Prime minister Erdogan's statement: ``No matter what you do, we took our final decision about Gezi Park." \\
		& A2 & 2013-06-02 & Erdogan refers to protesters as marauders (\emph{\c{c}apulcu}). \\
		& A3 & 2013-06-03 & Erdogan says ``There is 50 percent, and we can barely keep them at home. But we have called on them to calm down" before his trip to Morocco. \\
	
	\hline
	\multirow{5}{*}{\rotatebox[origin=c]{90}{Police\quad}}
		& B1 & 2013-05-30 & Police forces raids Gezi Park by using tear gas and destroys tents of protesters without any notice. \\
		& B2 & 2013-06-03 & Official statements about the first death and many injuries all around Turkey. \\
		& B3 & 2013-06-11 & Riot police enters Taksim square with water cannons and uses tear gas against the protesters. \\
		& B4 & 2013-06-15 & Police clears Gezi Park and takes out the protesters. Police starts to stake out Gezi Park.\\
	
	\hline
	\multirow{1}{*}{\rotatebox[origin=c]{90}{Protests\quad}}
		& C1 & 2013-06-04 & A library is built by the protesters in Gezi Park. \\
		& C2 & 2013-06-13 & Mothers join protests after Huseyin Mutlu's (Governor of Istanbul) calls to mothers to bring their children home. \\
		& C3 & 2013-06-17 & Silent protest in Taksim square held by a standing man. Many others gather after his protest. \\
	
	\hline
	\end{tabular}
\label{tab:events}
\end{table*}

In this work we focus on the Gezi Park protest, a social uprising whose events unfolded during May and June 2013 in Turkey. Political and policy issues related to this movement have been recently discussed in the social science literature \cite{gole2013gezi,kuymulu2013reclaiming}. Here instead we present an empirical analysis of the conversation about Gezi Park that occurred on Twitter. Our goal is to gain insight to the protest discussion dynamics.
In particular, we aim at exploring three different aspects of this conversation: \emph{(i)} its spatio-temporal dimension, to determine whether it was concentrated only in the country of inception, or if it acquired significant attention worldwide, and to assess how it started and what trends it generated; \emph{(ii)} what roles individuals played in this conversation and what influence they had on others, and whether such roles changed over time as information was diffused and the protests unfolded; \emph{(iii)} and how the online behavior of individuals changed over time in response to real-world events. 
To the best of our knowledge, this is the first study to explore the temporal evolution of online user roles and behaviors as a reflection of on-the-ground events during a social upheaval. We do so by means of computational tools and data-driven analyses.

\subsection*{Contribution and outline}

\begin{itemize}
\item We present methods to extract topically focused conversations about the social uprising surrounding Gezi Park and related trending topics of conversation on Twitter. (See \S~\ref{sec:data}.)

\item We explore the spatio-temporal characteristics of the conversation; that is, where tweets about Gezi Park originated and what locations shared the most similar topics and trends. This analysis yields clusters of cities that are mostly consistent with the country's geopolitics. (See \S~\ref{sub:spatio-temporal}.)

\item We analyze the emerging characteristics of users involved in the conversation about Gezi Park protests on Twitter, the roles they played in this context, and how these roles evolved as the protest unfolded. We find that influence was redistributed in the user population over time, making the conversation more democratic. (See \S~\ref{sub:roles}.)

\item We show that online user behavior was affected by external factors, such as speeches by political leaders or police action to hinder or suffocate the protests. (See \S~\ref{sub:external}.)
\end{itemize}

\begin{table*}[t]
\label{table:hashtags}
\caption{Set of hashtags commonly used by protesters and government supporters.}
\centering
\begin{tabular}{ l l l l }
	\hline
	Commonly used hashtags & & Local protest hashtags & Government supporters' hashtags \\
	\hline
	\hline
	\#direngeziparki &  \#bizeheryertaksim & \#direnankara & \#dunyaliderierdogan \\
	\#occupygezi & \#gezideyim & \#direnbesiktas & \#seviyoruzsenierdogan \\
	\#eylemvakti & \#7den77yedireniyoruz & \#direnizmir & \#seninleyizerdogan \\
	\#occupyturkey & \#heryertaksimheryerdirenis & \#direntaksim & \#seninleyiztayyiperdogan \\
	\#direngezi & \#korkakmedya & \#direnadana & \#youcantstopturkishsuccess \\
	\#tayyipistifa & \#hukumetistifa & \#direndersim & \#weareerdogan \\
	\#bubirsivildirenis & \#dictatorerdogan & \#direnistanbul & \#yedirmeyiz \\
	\#wearegezi & \#siddetidurdurun & \#direnrize & \#turkiyebasbakanininyaninda \\
	\hline
\end{tabular}
\label{tab:hashtags}
\end{table*}

\begin{table*}[t]
\caption{Trends in Turkey (country level) and in 12 Turkish cities during the observation period.}
\centering
	\begin{tabular}{lp{13cm}} 
	\hline
	Trend Location & Top 5 trending hashtags/phrases \\
	\hline
	Turkey & Turkey, Necati \c{S}a\c{s}maz, \#DirenGeziSeninleyiz, \#OyunaGelmiyoruzTakiple\c{s}iyoruz, \#Provokat\"orlereUYMA \\
	
	Istanbul & Turkey, Necati \c{S}a\c{s}maz, \#DirenGeziSeninleyiz, Bruno Alves, \#OyunaGelmiyoruzTakiple\c{s}iyoruz  \\
	
	Ankara & Turkey, Necati \c{S}a\c{s}maz, Bruno Alves, \#DirenGeziSeninleyiz, \#Provokat\"orlereUYMA\\
	
	Izmir & Turkey, Necati \c{S}a\c{s}maz, \#DirenGeziSeninleyiz, \#Tatil\"oncesiTakiple\c{s}elim, \#Provokat\"orlereUYMA  \\
	
	Bursa & Turkey, \#Tatil\"oncesiTakiple\c{s}elim, Necati \c{S}a\c{s}maz, \#K{\i}zlarTakiple\c{s}iyor, \#\c{c}apulcularTakiple\c{s}irse \\
	
	Adana & Turkey, \#\c{c}apulcularTakiple\c{s}irse, \#Tatil\"oncesiTakiple\c{s}elim, Necati \c{S}a\c{s}maz, \#DirenGeziSeninleyiz \\
	
	Gaziantep & Turkey, Necati \c{S}a\c{s}maz, \#SesVerT\"urkiyeBu\"UlkeSahipsizDe\u{g}il, \#DirenGeziSeninleyiz, \#OyunaGelmiyoruzTakiple\c{s}iyoruz \\
	
	Konya & \#Tatil\"oncesiTakiple\c{s}elim, Turkey, \#BizimDelilerTakiple\c{s}iyor, Necati \c{S}a\c{s}maz, \#SesVerT\"urkiyeBu\"UlkeSahipsizDe\u{g}il\\
	
	Antalya & Turkey, \#K{\i}zlarTakiple\c{s}iyor, \#CapulchularTakiple\c{s}iyor, \#T\"urkiyeBa\c{s}bakan{\i}n{\i}nYan{\i}nda, Necati \c{S}a\c{s}maz \\
	
	Diyarbakir & Turkey, Necati \c{S}a\c{s}maz, \#DirenGeziSeninleyiz, \#OyunaGelmiyoruzTakiple\c{s}iyoruz, \#Provokat\"orlereUYMA \\
	
	Mersin & \#HayranGruplar{\i}Takiple\c{s}iyor, Turkey, \#Tatil\"oncesiTakiple\c{s}elim, \#T\"urkiyemDireniyor, \#direnankara\\
	
	Kayseri & Turkey, Necati \c{S}a\c{s}maz, \#DirenGeziSeninleyiz, \#Seni\_G\"or\"unce, \#Provokat\"orlereUYMA \\
	
	Eskisehir & Turkey, Necati \c{S}a\c{s}maz, \#DirenGeziSeninleyiz, \#OyunaGelmiyoruzTakiple\c{s}iyoruz, \#Provokat\"orlereUYMA \\
	\hline
	\end{tabular}
\label{tab:trends}
\end{table*}

\section{Background about the protest} 
\label{sec:background}

In this section we provide some background information about the Gezi Park movement, explaining the context of the protests, the triggers for the mobilization, the timeline of events, and the ways which those events unfolded.

The protests began quietly in an already politically divided Turkey on May 28, 2013 with about 50--100 environmental activists who gathered for a sit-in at Gezi Park in Taksim Square, Istambul.  They were there to demonstrate against the destruction of one of the last public green spaces in central Istanbul.  The government had slated the space for the construction of a replica of  an Ottoman-era barracks that would be the site of luxury residences and a shopping mall.  The peaceful encampment  successfully resisted the demolition of the park by bulldozers when demonstrators refused to leave.  At dawn on the morning of May 30,  and then again the next morning, the protesters were attacked by the police using tear gas and water cannons, triggering clashes between authorities and the demonstrators that lasted until the end of the park occupation on June 15. During that time period,  the size of the groups of demonstrators escalated to about 10,000 on both the European and Asian sides of the Bosphorus and many thousands more in major cities across the country.  The focus of the protests grew from upset over Gezi Park's potential destruction to widespread criticism of the government's increasingly authoritarian practices and intrusions into the private lives of its citizens.  As the New York Times reported,
\begin{quote} In full public view, a long struggle over urban spaces is erupting as a broader fight over Turkish identity, where difficult issues of religion, social class and politics intersect. \cite{arango2013protests} \end{quote} 
Throughout the struggle, the protesters, who mostly consisted of middle-class secular Turks but also included some members of left-wing groups and nationalists, used social media to  alert others to their plans, urge others to join them, warn participants of police attacks and potential danger spots, provide information about makeshift medical assistance locations, and announce their goals.  A poll of about 3,000 activists found that the motivation of the demonstrators was their anger with Prime Minister Erdogan and not his political party or his aides.  More than 90\% of the respondents said they took to the streets because of Erdogan's authoritarian attitude \cite{uras2013what}.

A detailed timeline of the Gezi Park protests' major events during this period is provided in Table \ref{tab:events}.

\begin{figure*}[!ht]
\centering
\includegraphics[width=2.1\columnwidth]{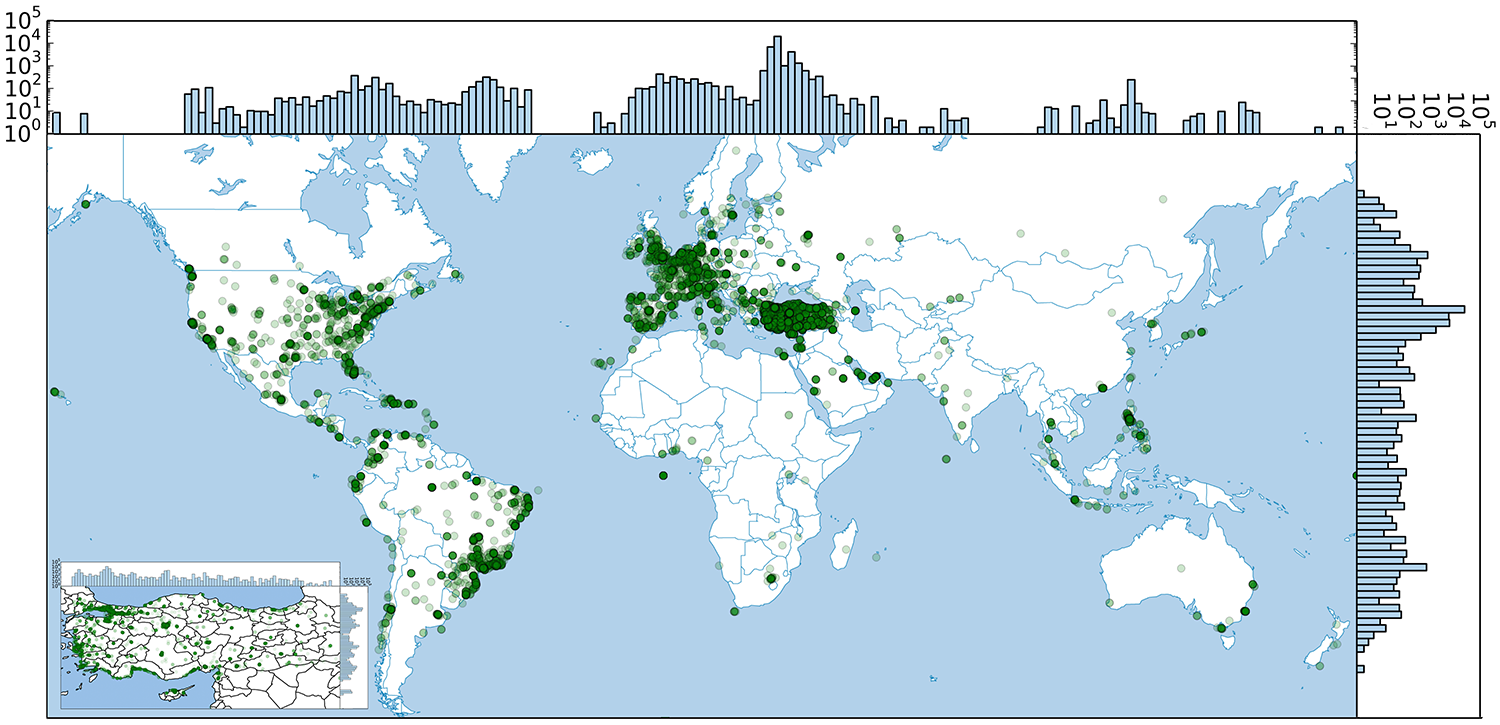}
\caption{Geographic distribution of tweets in our sample related to the discussion of Gezi Park events. The histograms represent the total volume by latitude and longitude. Content production crossed the Turkish national boundaries and spread in Europe, North and South America.}
\label{fig:gps_positioning_cropped}
\end{figure*}

\section{Data collection} 
\label{sec:data}

Our analysis is based on data collected from Twitter.
Twitter users can post \emph{tweets} up to 140 characters in length, which might contain URLs and media alongside text. Users can also interact with each other through various means, including the creation of directed social links (follower/followee relations), \emph{retweeting}  content (\emph{i.e.,} rebroadcasting messages to their followers), and \emph{mentioning} other users in their posts.
Tweets may also contain \emph{hashtags}, that are keywords used to give a topical connotation to the tweets (like \#direngeziparki and \#occupygezi). Multiple hashtags might co-occur in the same tweet. 

The dataset collected for our study comes from a 10\% random sample of all tweets streamed in real time, which was stored, post-processed and analyzed in-house. 
The observation period covers 27 days, from May 25th to June 20th, 2013: this time window started four days prior to the beginning of the Gezi Park events, and fully covered the three weeks during which the main protests unfolded. The short period prior to the protest inception is used as baseline to define user activity and interests. 

Our sample not only contains information about the tweets, but also meta-data about the users, including their \emph{screen names}, follower/followee counts, self-reported locations, and more. Additionally, for content posted with a GPS-enabled smartphone, we have access to the  geographic location from which the tweets were generated.

To isolate a representative sample of topical discussion about Gezi Park events, we adopted a hashtag seed-expansion procedure \cite{conover2011political}: first, we hand-picked the most popular Gezi Park related hashtag (\#diregeziparki) and we extracted all tweets containing this hashtag during our 27-day long period of interest. We then built the hashtag co-occurrence list, and we selected the top 100 hashtags co-occurring with our seed (\#diregeziparki). We generated our final list of hashtags of interest to include the set of commonly co-occurring hashtags and expanded our dataset collecting all tweets containing any of these hashtags. These hashtags were manually divided in three categories: general-interest hashtags, local protest related ones, and finally those used by government supporters. A detailed list containing the top general-purpose, local-protest, and government-support hashtags are listed in Table~\ref{tab:hashtags}.

Overall, we collected 2,361,335 tweets associated with the Gezi Park movement, generated by 855,616 distinct users and containing a total of 64,668 unique hashtags. Among these 2.3 million tweets, 1,475,494 are retweets and 47,163 are replies from one users to another. Also, 43,646 tweets have latitude/longitude coordinates. We adopt this subset of geolocated tweets to study the spatio-temporal nature of the protest (see \S~\ref{sub:spatio-temporal}).

During the same 27-day long observation period, we monitored the Twitter trends occurring at the country level in Turkey, and at the metropolitan area level in 12 major cities as provided by Twitter, namely: Adana, Bursa, Istanbul, Izmir, Kayseri, Gaziantep, Diyarbakir, Eskisehir, Antalya, Konya and Mersin. The list of top 10 hashtags and phrases trending both at the country level and at the city level were pulled from the platform at regular intervals of 10 minutes. This method \cite{ferrara2013traveling} is used in our analysis to define the similarity of topical interests and the patterns of collective attention towards Gezi Park conversation in the country. During this period we also monitored worldwide trends to determine whether and when the discussion about the protest achieved global visibility. A detailed list of the top popular trending hashtags and phrases for each location and at the country level is provided in Table \ref{tab:trends}.

\section{Results} \label{sec:results}

In this section we present the results of our analyses on spatio-temporal characteristics of the Gezi Park conversation on Twitter, evolving roles of the users involved, and effects of real-world events on online behaviors.


\subsection{Spatio-temporal cues of the conversation} 
\label{sub:spatio-temporal}

Our first analysis aims at determining the extent to which the discussion about Gezi Park attracted individual attention inside the national boundaries of Turkey, where the movement began, and how much of this conversation spread worldwide.

\begin{figure}[h]
\centering
\includegraphics[width=\columnwidth]{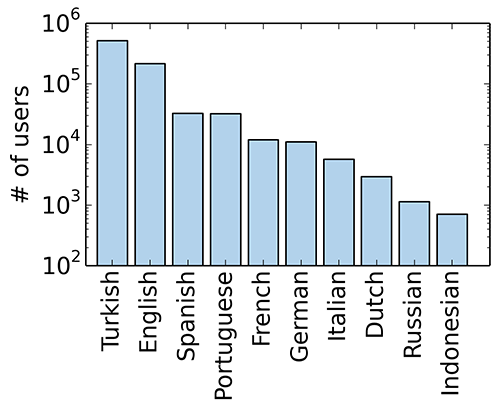}
\caption{Distribution of top 10 languages in tweets about the protest. Language information was extracted from the tweet meta-data.}
\label{fig:userlanguage_histogram}
\end{figure}

\begin{figure*}[t]
\centering
\includegraphics[width=.68\columnwidth]{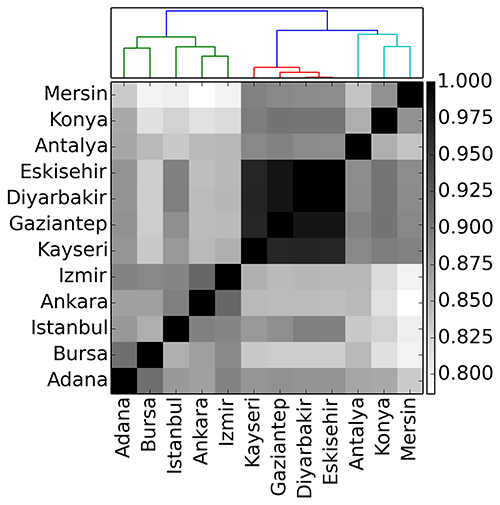}
\fbox{\includegraphics[width=1.32\columnwidth]{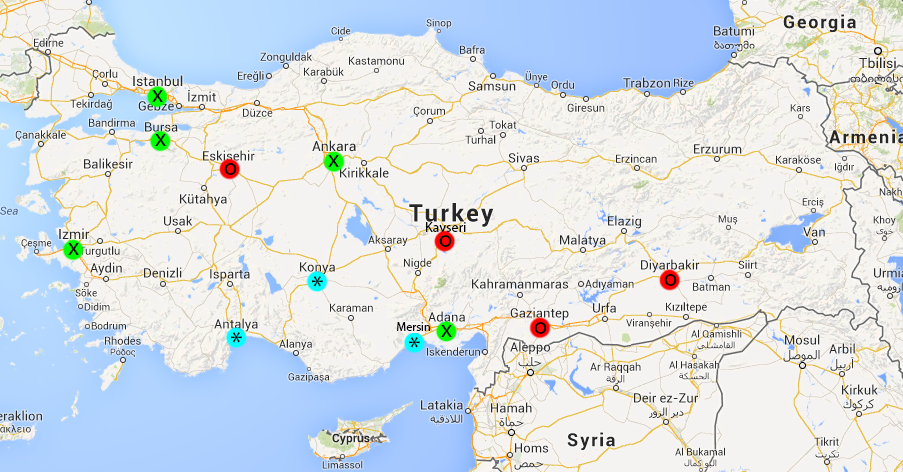}}
\caption{Left: Trend similarity matrix for 12 cities in Turkey. From the dendrogram on top we can isolate three distinct clusters. Right: Location of the cities with trend information, labeled by the three clusters induced by trend similarity.}
\label{fig:trend-similarity}
\end{figure*}

We focus on the subset of tweets in our dataset that have geo-coordinates attached (in the form of latitude/longitude). Such tweets are likely to be posted by GPS-enabled devices (like smartphones) and represent only a small fraction of total tweets ($\approx 1.84\%$ of our sample), which is consistent with similar studies \cite{conover2013geospatial}. Yet they provide a very precise picture of the geospatial dynamics of content production. Figure \ref{fig:gps_positioning_cropped} maps the sources of these tweets. The figure also shows histograms on the horizontal and vertical axes, that illustrate the distribution of tweets occurring in the corresponding locations, binned by latitude and longitude. From this figure the global nature of the discussion about Gezi Park events clearly emerges. Although a large fraction of tweets originated in Turkey, a significant amount was produced in Europe, North and South America (especially the United States and Brazil). Other noteworthy countries involved in the discussion are the Philippines, Bahrain, Qatar and the United Arab Emirates. 

Attention abroad was signaled by the presence of trending hashtags and phrases in the worldwide Twitter trends. Among these, the main protest hashtag, \#direngeziparki, trended several times between May 31st and June 2nd, 2013; \#TayipIstifa, invoking Erdogan's resignation, appeared on June 6th, 2013. Worldwide attention is also evident in the variety of languages exhibiting hashtags related to the Gezi Park events, as displayed in Figure~\ref{fig:userlanguage_histogram}. After Turkish, the most popular languages were English, Spanish, and Portuguese. 

We also explored the local dimension of the conversation, focusing on the discussion inside the Turkish borders. Our goal was to determine whether any patterns of discussion of similar topics of conversation emerged. In Figure \ref{fig:trend-similarity} we show the trend similarity matrix computed among the sets of trending hashtags and phrases occurring in each of the 12 cities where Twitter trends are monitored. 
Each location is described by a frequency vector of occurrences of the observed trends. The similarity between pairs of cities is calculated as the cosine similarity of their trends frequency vectors.
Above the matrix we show the dendrogram produced by hierarchical clustering, where it is possible to appreciate the separation in three clusters. Such clusters neatly correspond to three different geographic areas of Turkey.  Physical proximity seems to play a crucial role in  determining the similarity of topical interests of individuals, consistent with other recent results \cite{ferrara2013traveling}.

The clusters found with our trend similarity analysis also seem to match the Turkish geopolitical profiles. Eskisehir, Kayseri and Gaziantep (in the red cluster) are all central Anatolian cities where the president's party (AKP) has a stronghold (though the CHP opposition party edged out the AKP in the March 2014 mayoral race); they are more culturally conservative and homogeneous. Izmir, Istanbul, Bursa, Ankara, and Adana (green cluster) are the largest cities in Turkey with diverse populations. Finally, Antalya and Mersin (blue cluster) are seacoast cities that are known for supporting the one of the main opposition parties (CHP or MHP). Further work is needed to understand why Konya is assigned to this cluster, as it is considered a major religiously conservative center (where the AKP mayoral candidate secured more than 64\% of the vote in the 2014 mayoral elections) that has little in common with the Mediterranean cities.

\begin{figure*}[t]
\centering
\includegraphics[width=2.1\columnwidth]{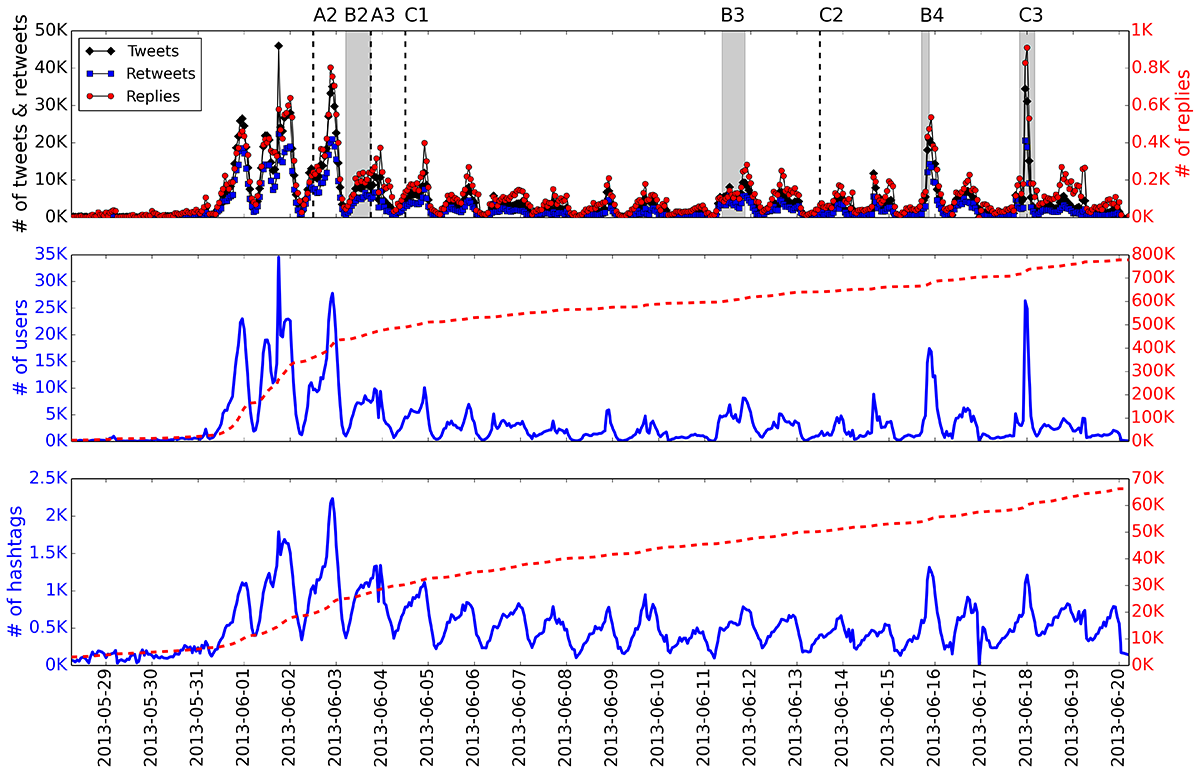}
\caption{Hourly volume of tweets, retweets and replies between May 30th and June 20th, 2013 (top). The timeline is annotated with events from Table \ref{tab:events}. User (center) and hashtag (bottom) hourly and cumulative volume of tweets over time.}
\label{fig:time_series_paper}
\end{figure*}

Let us explore the temporal dimension of the Gezi Park discussion. We wanted to determine whether the activity on social media mirrored on-the-ground events, and whether bursts of online attention coincided with real-world protest actions. We analyzed the time series of the volume of tweets, retweets and replies occurring during the 27-day-long observation window, as reported in Figure \ref{fig:time_series_paper} (top panel). 
The discussion was driven by bursts of attention that largely corresponded to major on-the-ground events (cf. Table \ref{tab:events}), similar to what has been observed during other social protests \cite{conover2013digital}. It is also worth noting that the numbers of tweets and retweets are comparable throughout the entire duration of the conversation, suggesting a balance between content production (\emph{i.e.,} writing novel posts) and consumption (\emph{i.e.,} reading and rebroadcasting posts via retweets). In the middle panel of Figure \ref{fig:time_series_paper} we report the number of users involved in the conversation at a given time, and the cumulative number of distinct users over time (dashed red line); similarly, in the bottom panel of the figure, we show the total number of hashtags related to Gezi Park observed at a given time, and the cumulative number of distinct hashtags over time.
We note that approximately 60\% of all users observed during the entire discussion joined in the very first few days, whereas additional hashtags emerged at a more regular pace throughout a longer period. This suggests that the conversation acquired traction immediately, and exploded when the first on-the-ground events and police action occurred.

\subsection{User roles and their evolution} 
\label{sub:roles}

Our second experiment aims at investigating what roles users played in the Gezi Park conversation and how they exercised their influence on others. We also seek to understand whether such roles changed over time, and, if so, to what extent such transformation reshaped the conversation.

\begin{figure}[t]
\centering
\includegraphics[width=\columnwidth]{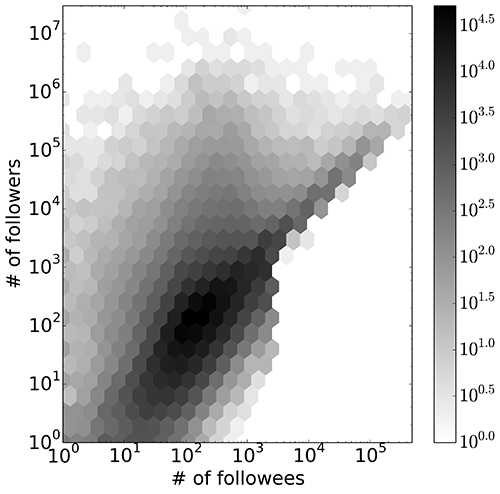}
\caption{Distribution of friends and followers of users involved in the Gezi Park conversation.}
\label{fig:friends-follower}
\end{figure}

Figure \ref{fig:friends-follower} shows the distribution of social ties reporting the two modalities of user connectivity, namely followers (incoming) and followees (outgoing) relations. The dark cells along the diagonal indicate that most users have a balanced ratio of ingoing and outgoing ties. Users below the diagonal follow more than they are followed. Note that most users are allowed to follow at most 1000 people. Finally, above the diagonal, we observe users with many followers. Note the presence of extremely popular users with hundreds of thousands or even millions of followers. The number of followers has a broad distributions and seems largely independent of the number of followees.

\begin{figure}[t]
\centering
\includegraphics[width=\columnwidth]{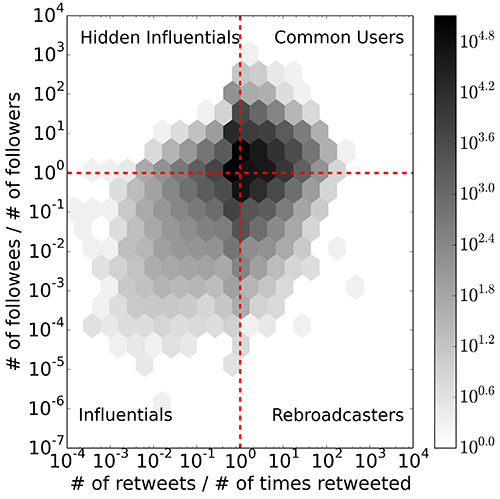}
\caption{Distribution of user roles as function of social ties and  interactions.}
\label{fig:influence_productivity}
\end{figure}

The presence of highly followed users in this conversation raises the question of whether their content is highly influential.
Following a methodology inspired by Gonz{\'a}lez-Bail{\'o}n \textit{et al.}~\cite{gonzalez2013broadcasters}, we determined user roles as a function of their social connectivity and interactions. 
Figure \ref{fig:influence_productivity} gives an aggregated picture of the distribution of user roles during the Gezi Park conversation. The y-axis shows the ratio between number of followees and followers of a given user; the x-axis shows the ratio between the number of retweets produced by a user and the number of times other users retweet that user. In other words, the vertical dimension represents social connectivity, whereas the horizontal dimension accounts for information diffusion. We can draw a vertical line to separate influential users on the left (\emph{i.e.,} those whose content is most often retweeted by others) and information consumers on the right (those who mostly retweet other people's content). Influential users can be further divided in two classes: those with more followers than followees (bottom-left) and those with fewer followers (top-left), which we call \emph{hidden influentials}. Similarly, information consumers can be divided in two groups--rebroadcasters with a large audience (bottom-right), and common users (top-right).

\begin{figure*}
\centering
\includegraphics[width=2.1\columnwidth]{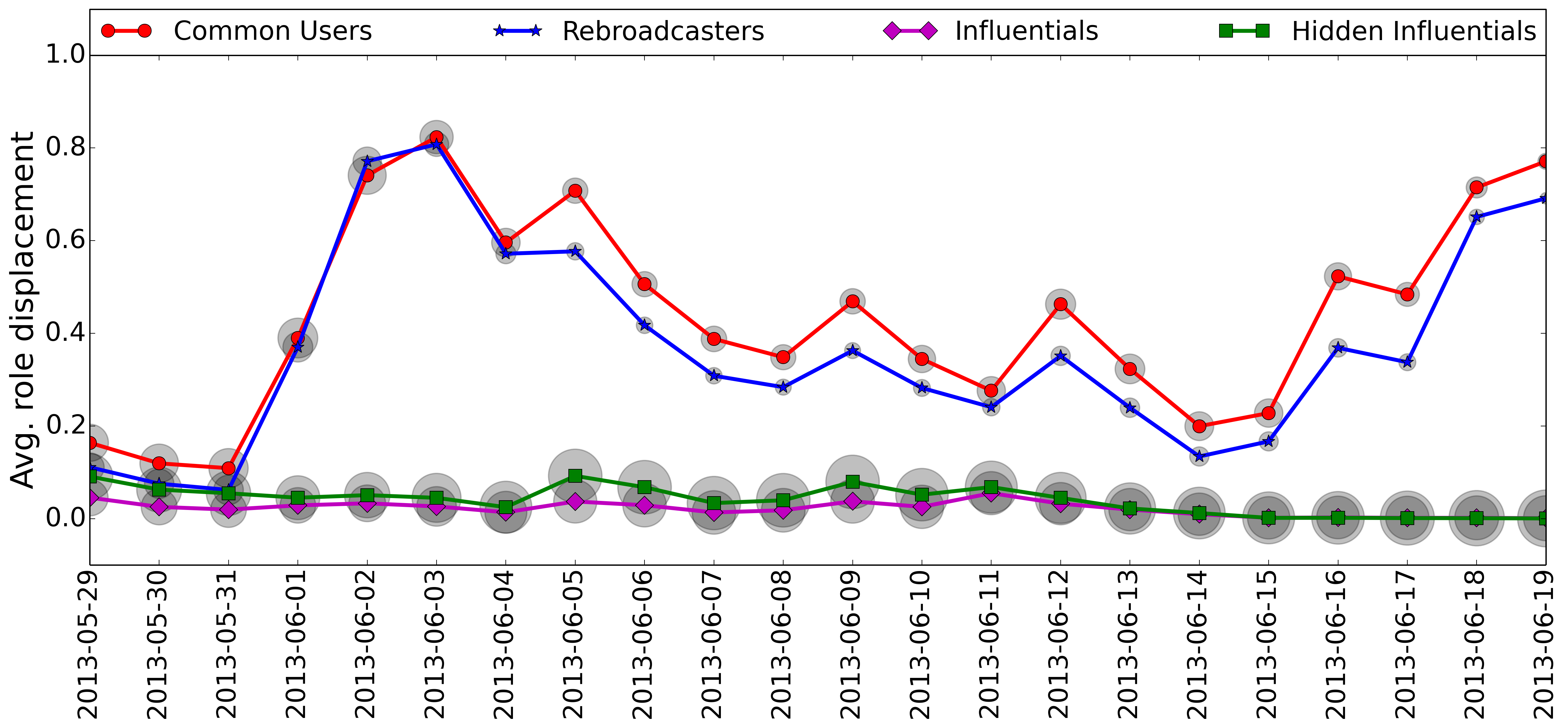}
\caption{Average displacement of roles over time for the four different classes of roles. The size of the circles represents the number of individuals in each role.}
\label{fig:roles-temporal}
\end{figure*}

Figure \ref{fig:influence_productivity} shows a static picture of aggregated data over the 27-day observation period. To study how roles evolve as events unfold, we carried out a longitudinal analysis whose results are provided in Figure \ref{fig:roles-temporal}. This figure shows the average displacement of each role class, and the number of individuals in each class (circles), for each day. The displacement is computed in the role space (that is, the space defined by the two dimensions of Figure \ref{fig:influence_productivity}). Larger displacements suggest that individuals in a class, on average, are   moving toward other roles. 

Various insights emerge from Figure~\ref{fig:roles-temporal}: first, we observed that the classes of information producers (influentials and  hidden influentials) are relatively stable over time; together they include more than 50\% of users every day, suggesting that many individuals in the conversation had large audiences, and the content they produced was heavily rebroadcasted by others (information consumers as well as other influentials). 
On the other hand, information consumers show strong fluctuation: starting from an initial configuration with stable roles (May 29--31),  common users and rebroadcasters subsequently exhibit large aggregate displacements in the role space (June 1--4). We also note a redistribution of the users in each role: at the beginning of the protest a large fraction represents common users and rebroadcasters, while, as time passed and events unfolded, these two classes shrank. This suggests that common users and rebroadcasters acquired visibility and influence over time: some fraction of these users moved from the role of information consumers to that of influentials, such that their content wass consumed and rebroadcasted by others. In other words, the discussion became more \emph{democratic} over time, in that the control of information production was redistributed to a larger population, and individuals acquired influence as the protests unfolded.

\subsection{Online behavior and exogenous factors} 
\label{sub:external}

\begin{figure}[t]
\centering
\includegraphics[width=\columnwidth]{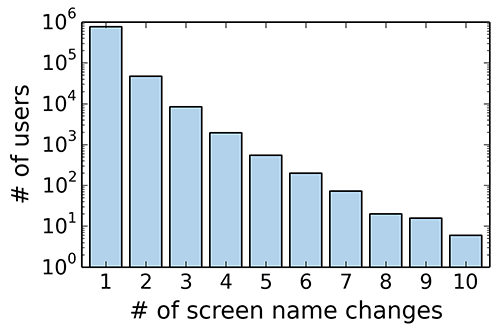}
\caption{Distribution of the number of screen name changes among users during the Gezi Park events.}
\label{fig:username_histogram}
\end{figure}

Our concluding analysis focused on the way on-the-ground events affected online user behavior. While analyzing our dataset we noticed an abnormal number of screen name changes, as reported in Figure~\ref{fig:username_histogram} (the screen name, not to be confused with the user name, is the name displayed in one's Twitter account). Many users changed their screen names five or more times. This was an unusual observation that attracted our attention. 

\begin{figure}[t]
\centering
\includegraphics[width=\columnwidth]{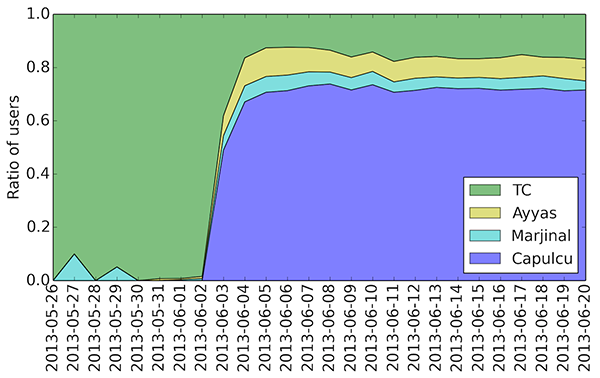}
\caption{Among the many users who changed screen names, this chart plots the fractions who adopted different nicknames over time in respons to external events.}
\label{fig:username_temporal}
\end{figure}

Further investigation revealed a collective synchronization process, as displayed in Figure \ref{fig:username_temporal}. The changes in screen names represent reactions of users involved in the Gezi Park conversation to external events: these users changed their Twitter screen names to reflect sobriquets attributed to them by their political leaders. One example is the adoption of ``TC'' (standing for \emph{Turkiye Cumhuriyeti} --- Turkish Republic). As a reaction to identity issues, several users started using TC in front of their screen names. Another relevant example is Erdogan's speech of June 2, during which he referred to protesters as marauders (\emph{\c{c}apulcu}), marginals (\emph{marjinal} or drunks (\emph{ayyas}). Individuals responded by changing their screen names to include such nicknames as a sign of protest against the government's attempt to discredit the protest participants and minimize the relevance of their actions. 
This phenomenon illustrates how online and offline worlds are tightly interconnected, deeply affecting each other.

\section{Related work}


The role of communication technologies used during social upheavals has been studied in the context of different events, including Arab Spring movements, Occupy Wall Street, and the Spanish 'Indignados' uprisings \cite{borge2011structural,choudhary2012social,conover2013geospatial,conover2013digital,gonzalez2011dynamics}. 
The benefits resulting from the adoption of social media include lowered barriers to participation, increased ease with which small-scale acts can be aggregated, the rapid propagation of logistic information and narrative frames, and a heightened sense of community and collective identity~\cite{bennett2003communicating,bennett2007changing,myers1994communication,van2009cyber,wray1999electronic}.

Gonz{\'a}lez-Bail{\'o}n \emph{et al.} \cite{gonzalez2011dynamics} collected a large corpus of tweets related to the Spanish social and economic `Indignados' protest that unfolded during May 2011. 
Their work provides evidence that Twitter played a role in the recruitment of new individuals to the protest movement as well as in the dissemination of information related to mass mobilization activities. 

Choudhary \emph{et al.} \cite{choudhary2012social} analyzed the aggregate tweet sentiment during the 2011 Egyptian revolution, observing that fluctuations in positive and negative sentiment were closely correlated with the sentiment expressed by influential users worldwide. The authors also observed that users tweeting about the Egyptian revolution were distributed both inside and outside Egypt.
Our work supports the global dimension of social protest discussion on social media: here we showed how Twitter brought worldwide visibility to the discussion of the Gezi Park protest, crossing the boundaries of its country of origin.

Recently, Ba{\~n}os \emph{et al.} \cite{banos2013role,banos2013diffusion} highlighted the role of social media users in the diffusion of information related to mass political mobilizations, unveiling the presence of hidden influentials who foster large cascades. The authors also observed how the topology of the communication network during such events reflects underlying dynamics like information diffusion and group emergence. Our analysis builds on this work and shows that user roles are not static, but rather evolve dynamically as the protest unfolds.  

In our recent work we studied the Occupy Wall Street uprising. First, we focused on the geospatial characteristic of the protest \cite{conover2013geospatial}. We observed that highly-localized discussions mirrored individuals' attempts to organize and coordinate mobilization on the ground. Interstate discussion channels driving long-distance communications fostered the collective framing process that imbues social movements with a shared language, purpose and identity.  A longitudinal analysis~\cite{conover2013digital} revealed that users did not change their connectivity, interests and attention patterns with respect to baseline activity prior to the beginning of the protest. These findings left open the question whether Occupy had any long-lasting effect on its online community of participants.

\section{Conclusions}

In this paper we focused on the analysis of the conversation about the Gezi Park protest that took place on Twitter. We collected a large dataset spanning the time from May 25th to June 20th, 2013, during which the main events of the protests unfolded. 

Our analysis of the spatial dynamics of the communication brought two different interesting findings. First, we observed that the discussion about Gezi Park events spread worldwide, and a sustained number of tweets was produced over time outside of Turkey --- in Europe, North and South America. International attention was underscored by trending hashtags related to Gezi Park at the worldwide level. Second, we observed that local trends followed geographic and political patterns. Among the 12 cities whose trends we monitored, three clear geographic clusters emerge. This result is consistent with our recent analysis of geospatial spreading patterns of Twitter trends \cite{ferrara2013traveling}. The discussion was driven by bursts of attention that largely correlated with on-the-ground events.

Focusing on users, we identified four types of roles (common users, rebroadcasters, influentials and hidden influentials). We tracked their evolution over time as events unfolded. As time passed, the discussion about Gezi Park became more democratic, with an increased number of influential users.

Our analysis concluded by studying an effect of real-world events, such as political speeches, on online user behavior. We found that individuals responded to such external provocations by exhibiting collective actions, namely the change of their Twitter screen names to reflect sobriquets attributed to them by their political leaders. 

Our analysis uncovered various interesting dynamics, yet much remains to be done. It would be interesting to investigate whether universal patterns of communication emerge from different classes of conversation, including those about social and political issues, if contrasted with  other types like sports, news, or entertainment. This would allow us to separate intrinsic characteristics of human communication dynamics from topic specific patterns.

\section*{Acknowledgments}
This work was partially supported by NSF (grant CCF-1101743), DARPA (grant W911NF-12-1-0037), and the McDonnell Foundation. The funders had no role in study design, data collection and analysis, decision to publish, or preparation of the manuscript.

\bibliographystyle{abbrv}
\bibliography{sigproc}  
%
%
\end{document}